\documentclass[letterpaper]{article}
\pdfoutput=1 

\usepackage{jheppub} 

\usepackage[T1]{fontenc} 

\newcommand{\psib}{{\overline{\psi}}}

\title{\boldmath Fermion masses through four-fermion condensates}

\author[a]{Venkitesh Ayyar}
\author[a,b]{Shailesh Chandrasekharan}
\affiliation[a]{Department of Physics, Duke University, Durham, NC 27708, USA}
\affiliation[b]{Center for High Energy Physics, Indian Institue of Science, Bangalore, 560 012, India}

\emailAdd{vpa@phy.duke.edu}
\emailAdd{sch@phy.duke.edu}

\abstract{Fermion masses can be generated through four-fermion condensates when symmetries prevent fermion bilinear condensates from forming. This less explored mechanism of fermion mass generation is responsible for making four reduced staggered lattice fermions massive at strong couplings in a lattice model with a local four-fermion coupling. The model has a massless fermion phase at weak couplings and a massive fermion phase at strong couplings. In particular there is no spontaneous symmetry breaking of any lattice symmetries in both these phases. Recently it was discovered that in three space-time dimensions there is a direct second order phase transition between the two phases. Here we study the same model in four space-time dimensions and find results consistent with the existence of a narrow intermediate phase with fermion bilinear condensates, that separates the two asymptotic phases by continuous phase transitions.}

\begin{document} 
\maketitle
\flushbottom

\section{Introduction}

Masses of free fermions arise from local fermion bilinear terms in the action. If symmetries of the theory prevent such terms, fermions remain massless perturbatively. However, these symmetries can break spontaneously and generate non-zero fermion bilinear condensates that can make fermions massive. This traditional mechanism of fermion mass generation is well known and is used in the standard model of particle physics to give quarks and leptons their masses. In QCD, along with confinement, this mechanism also helps explain the existence of light pions while making nucleons heavy. In this work we explore a different mechanism of fermion mass generation where fermions acquire their mass through four-fermion condensates, while fermion bilinear condensates vanish. This alternate mechanism has been the focus of many recent studies in 3D lattice models \cite{Slagle:2014vma,He:2015bda,Ayyar:2014eua,Catterall:2015zua,Ayyar:2015lrd,He:2016sbs}. Here we explore if these results extend to 4D. The 3D studies also show that no spontaneous symmetry breaking of any lattice symmetries is necessary for fermions to become massive. The presence of a new second order critical point makes the mechanism interesting even in the continuum.\footnote{The fermion mass generation mechanism we explore in this work is different from the one proposed in \cite{Stern:1998dy,PhysRevD.59.016001,Kanazawa:2015kca} where chiral symmetry is spontaneously broken due to four-fermion condensates instead of fermion bilinear condensates and massless bosons are present. In the massive fermion phase we explore here all particles, including bosons are massive.}

We believe that the alternate mechanism of mass generation can be understood qualitatively if we view the four-fermion condensate as a fermion bilinear condensate between a composite fermion (consisting of three fundamental fermions) and a fundamental fermion. When three fundamental fermions bind to form a composite fermion state, the four-fermion condensate can begin to act like a conventional mass term. However, since such composite states can only form at sufficiently strong couplings a non-perturbative approach is required to uncover it. At weak couplings, when composite states do not form, four-fermion condensates cannot act like mass terms although they are still non-zero. Since there are no local order parameters that signal the formation of the composite fermion bound states, the massive phase does not require spontaneous symmetry breaking. All these arguments are consistent with the results in 3D lattice models mentioned above.

Generating fermion masses through interactions but without spontaneous symmetry breaking is a subtle problem from the perspective of 4D continuum quantum field theories due to anomaly matching arguments \cite{tHooft:1979bh,PhysRevLett.45.100,Banks:1991sh,Banks:1992af}, but 4D lattice models that display such a mechanism of mass generation are well known and have been studied extensively in the context of lattice Yukawa models with both staggered fermions \cite{Hasenfratz:1988vc,Hasenfratz:1989jr,Lee:1989mi} and Wilson fermions \cite{Bock:1990tv,Bock:1990cx,Golterman:1990yb,Gerhold:2007gx,Bulava:2011jp}. These models contain a massless fermion phase at weak couplings (referred to as the paramagnetic weak or PMW phase), and a non-traditional symmetric massive fermion phase at strong couplings (referred to as the paramagnetic strong or PMS phase). The fermion mass in the PMS phase can be argued as being generated due to four fermion condensates since fermion bilinear condensates vanish in that phase. A review of the early work can be found in \cite{Shigemitsu:1991tc}.

In order for the PMS phase found in previous lattice calculations to become interesting from the point of view of continuum quantum field theory, it must be possible to tune the fermion mass to zero in lattice units. In units where the fermion mass remains fixed, this would imply that the lattice spacing vanishes. This can be accomplished in the presence of a direct second order transition between the PMW and the PMS phase. Such a transition was proposed as an important ingredient for realizing chiral fermions on the lattice \cite{Eichten:1985ft,Golterman:1992yha}. Unfortunately, all previous studies found that there was always an intermediate phase (referred to as the ferromagnetic of FM phase) where the symmetry that protected the fermions from becoming massive at weak couplings, was broken spontaneously. In the presence of the FM phase, fermions in the PMS phase cannot be made arbitrarily light in lattice units. We believe this was the reason the PMS phase was abandoned as merely a lattice artifact. In fact earlier studies in 3D also found an intermediate FM phase \cite{Alonso:1999hh}. Hence the recent discovery of a possible direct second order PMW-PMS phase transition in 3D is exciting, and raises the possibility that such transitions may exist even in 4D. 

A second order PMW-PMS transition does not fall under the usual Landau-Ginzburg paradigm, since both the phases have the same global symmetries and there is no local order parameter that distinguishes them. For this reason it must be different from the usual Gross Neveu universality class. Such transitions are known in condensed matter literature and usually driven due to a change in the topological properties of the ground state \cite{Senthil:2014ooa}.  The PMW-PMS transition could occur due to a similar reason although it does not seem to involve any topological order \cite{He:2016sbs}. From a condensed matter perspective, the PMS phase can be viewed as a trivial insulator where the ground state does not break any lattice symmetries since it is formed by local singlets. In contrast, the traditional massive fermion phase with fermion bilinear condensates is like a gapped semi-metal or a topological insulator. Topological insulators can have chiral zero modes attached to domain walls where the sign of the condensate changes \cite{Callan:1984sa}. Such zero modes are extensively used today in lattice QCD studies through the domain wall formulation introduced by Kaplan \cite{Kaplan:1992bt}. Many interesting properties of such topological insulators in background fields have also been studied by particle physicists many years ago \cite{Golterman:1992ub,Kaplan:1999jn}. Recently the focus has shifted to the classification of the topological insulators in the presence of fermion self interactions \cite{Fidkowski:2009dba,PhysRevB.83.075103,1367-2630-15-6-065002,Fidkowski:2013jua,PhysRevB.92.125104,PhysRevB.89.195124}. These studies suggest that when the fermion content of the theory is chosen appropriately, such interactions can smoothly deform a topological insulator to a trivial insulator. During such a change massless chiral fermions on the edges acquire non-traditional masses due to the formation of four-fermion condensates since fermion bilinear condensates are forbidden. The associated phase transition on the edge need not involve any spontaneous symmetry breaking \cite{Slagle:2014vma}. This has prompted many applications of the alternate mass generation mechanism to particle physics \cite{Wen:2013ppa,You:2014vea,You:2014oaa,BenTov:2015gra}.

A direct second order PMW-PMS phase transition has remained elusive in 4D so far. Given the discovery of such a transition in 3D \cite{Ayyar:2015lrd}, we believe it is worth searching for it even in 4D. The first step obviously would be to explore if the same lattice model that showed its presence in 3D, also contains it in 4D. Interestingly, this model contains sixteen Weyl fermions, which has been argued to be the right number necessary for the possible existence of the transition in 4D \cite{You:2014vea}. However, this model was already studied long ago in the context of Higgs-Yukawa models and a {\em wide} intermediate FM phase was found \cite{Lee:1989mi}, implying that one needs to explore extensions to it. It may be possible to add new couplings to the model that have the effect of narrowing the width of the FM phase. Unfortunately, the conclusions of the earlier work were mostly drawn from mean field theory and crude Monte Carlo calculations. Hence, in this work we focus on accurately determining the phase boundaries of the model so as to get a sense of how far away is the possible critical point in the extended parameter space. By working in the limit where the Higgs field can be integrated out explicitly, we can accurately study the model in the chiral limit with Monte Carlo methods on lattices up to $12^4$ using the fermion bag approach \cite{PhysRevD.82.025007,Chandraepja13}. In contrast to the earlier work, our results shows a surprisingly {\em narrow} intermediate FM phase, assuming it exists. 

Our paper is organized as follows. In section \ref{sec2} we provide a new view point for our lattice model and discuss its symmetries. We also discuss observables that shed light on the phase structure of the model. In section \ref{sec3} we present the fermion bag approach and show that fermion bags have interesting topological properties. In particular we discuss an index theorem very similar to the one in non-Abelian gauge theories with massless fermions. In section \ref{sec4} we explain how the fermion bag approach provides a new theoretical perspective on the physics of the PMS phase and the alternate mass generation mechanism. In particular we explain how all fermion bilinear mass order parameters in the model must vanish at sufficiently strong couplings, although fermions are massive. In section \ref{sec5} we present our Monte Carlo results and in section \ref{sec6} we present our conclusions.

\section{The Model} 
\label{sec2}

Our model was originally studied within the context of lattice Higgs-Yukawa models \cite{Lee:1989mi}. However, it can also be obtained directly by discretizing naively the continuum four-fermion action containing a single Dirac fermion field $\psi^a(x)$ and $\psib^a(x)$ where $a$ labels the four spinor indices. We believe this alternate view point sheds more light on the mechanism of mass generation with four-fermion condensates (or equivalently the PMS phase) at strong couplings. Consider the continuum Euclidean action given by
\begin{equation}
S_{\rm cont} = \int d^4x \ \Big\{\ \psib(x)\gamma_\alpha\partial_\alpha \psi(x) \ -\  U\Big(\psi^4(x) \psi^3(x) \psi^2(x) \psi^1(x) + \psib^4(x) \psib^3(x) \psib^2(x) \psib^1(x) 
\Big) \Big\}.
\label{contact}
\end{equation}
where $\gamma_\alpha$ are the usual $4 \times 4$ Hermitian Dirac matrices. Note that the continuum model breaks the $U(1)$ fermion number symmetry 
\begin{equation}
\psi(x) \rightarrow \exp(i\theta)\psi(x), \ \ \psib(x) \rightarrow \exp(-i\theta)\psib(x)
\label{u1fs}
\end{equation} 
explicitly, but it is invariant under Euclidean rotations and the $U(1)$ chiral symmetry 
\begin{equation}
\psi(x) \rightarrow \exp(i\theta \gamma_5)\psi(x), \ \ \psib(x) \rightarrow 
\psib(x)\exp(i\theta \gamma_5).
\label{u1cs}
\end{equation} 
Perturbatively, no fermion bilinear mass term can be generated through radiative corrections since all such terms break either the $U(1)$ chiral symmetry or the rotational symmetry. Thus, the model must contain a massless fermion phase (or the PMW phase) at weak couplings. At strong couplings, assuming we can perform a perturbative (strong coupling) expansion in the kinetic term (similar to the hopping parameter expansion on the lattice), we can see the presence of a symmetric massive fermion phase (or the PMS phase).
In particular the leading order theory is trivial since all fermion fields are bound into local space-time singlets under the symmetries of the action. Introduction of the kinetic term can create excitations that transform non-trivially under both chiral and rotational symmetries, but all of these must be massive since energetically favored singlets need to be broken to create them. But, can the strong coupling expansion as described above be justified after the subtleties of UV divergences are taken into account? Although we cannot answer this question for a single Dirac field, ignoring the fermion doubling problem, we can easily discretize the continuum action (\ref{contact}) naively on the lattice and ask the same question in a controlled setting in the lattice theory. In particular we can even explore if the fermion mass of the lattice theory at strong couplings (i.e., in the PMS phase) can be made light as compared to the cutoff. 

Discretizing (\ref{contact}) naively on a space-time lattice we obtain
\begin{equation}
S_{\rm naive} = \sum_{x,y} \ \Big\{\ \psib_x \gamma_\alpha \frac{1}{2}(\delta_{x+\hat{\alpha},y} - \delta_{x-\hat{\alpha},y})\psi_y\ -\  U \Big(\psi^4_x \psi^3_x \psi^2_x \psi^1_x + \psib^4_x \psib^3_x \psib^2_x \psib^1_x\Big) \Big\},
\end{equation}
where we use the notation $\psi^a_x$ to denote the lattice Grassmann fields. Using the well known spin diagonalization transformation
\begin{equation}
\psi_x \rightarrow (\gamma_1)^{x_1}(\gamma_2)^{x_2}(\gamma_3)^{x_3}(\gamma_4)^{x_4} \psi_x,\ \ 
\psib_x \rightarrow \psib_x (\gamma_4)^{x_4}(\gamma_3)^{x_3}(\gamma_2)^{x_2}(\gamma_1)^{x_1} ,\ \
\end{equation}
used to define staggered fermions \cite{Sharatchandra:1981si}, we obtain the lattice action,
\begin{equation}
S_{\rm naive} = \sum_{x,y} \ \Big\{\ \psib_x M_{x,y}\psi_y\ -\  U\Big(\psi^4_x \psi^3_x \psi^2_x \psi^1_x + \psib^4_x \psib^3_x \psib^2_x \psib^1_x\Big) \Big\}.
\end{equation}
where $M_{x,y}$ is the free staggered fermion matrix 
\begin{equation}
M_{x,y} \ =\ \frac{1}{2}\ \sum_{\alpha} \ \eta_{\alpha,x}\ \big(\delta_{x+\hat{\alpha},y} -  \delta_{x-\hat{\alpha},y}\big).
\end{equation}
The phase factors $\eta_{1,x}=1$, $\eta_{2,x}=(-1)^{x_1}$, $\eta_{3,x}=(-1)^{x_1+x_2}$, $\eta_{4,x}=(-1)^{x_1+x_2+x_3}$ are well known. Since $\psi^a_x$ on even sites only connect with $\psib^a_x$ on odd sites and vice versa, we can eliminate half the degrees of freedom by defining $\psi^a_x$ only on even sites and $\psib^a_x$ only odd sites. We can go a step further and stop distinguishing between $\psi^a_x$ and $\psib^a_x$ since every site has an identical single four component Grassmann variable. This finally leads to the Euclidean action,
\begin{equation}
S \ =\ \frac{1}{2}\sum_{x,y,a}\ \psi^a_x \ M_{x,y} \ \psi^a_y \  - \ U\ \sum_x \ \psi^4_x \psi^3_x \psi^2_x \psi^1_x.
\label{act}
\end{equation}
We can also view the above action as being constructed directly with four reduced flavors of staggered fermions with an onsite four-fermion interaction. In this interpretation, the spinor indices $a=1,2,3,4$ are viewed as labels of the four reduced staggered flavors. When $U=0$ the above model describes eight flavors of Dirac fermions (or equivalently sixteen flavors of Weyl fermions) in the continuum. This matches the required number of fermions that allows for a non-traditional massive phase according to recent insights \cite{You:2014vea}. 

Lattice symmetries of staggered fermions are well known \cite{Golterman:1984cy}. These include: \\
\noindent (1) {\em Shift Symmetry:}
\begin{equation}
\psi^a_x \rightarrow \xi_{\rho,x} \psi^a_{x+\rho},
\end{equation}
where $\xi_{1,x}=(-1)^{x_2+x_3+x_4}$, $\xi_{2,x}=(-1)^{x_3+x_4}$, $\xi_{3,x}=(-1)^{x_4}$, $\xi_{4,x}=1$. This symmetry is based on the identity $\xi_{\rho,x} \eta_{\alpha,x}\xi_{\rho,x+\hat{\alpha}} = \xi_{\rho,x} \eta_{\alpha,x}\xi_{\rho,x-\hat{\alpha}} = \eta_{\alpha,x+\hat{\rho}}$.\\
\noindent (2) {\em Rotational Symmetry:}
\begin{equation}
\psi^a_x \rightarrow S_R(R^{-1}\ x)\psi^a_{R^{-1}\ x},
\end{equation}
where $R \equiv R^{(\rho\sigma)}$ is the rotation $x_\rho \rightarrow x_\sigma$, $x_\sigma = -x_\rho$, $x_\tau \rightarrow x_\tau$ for $\tau \neq \rho,\sigma$ and
\begin{equation}
S_R(x) = \frac{1}{2}[1\pm \eta_{\rho,x} \eta_{\sigma,x} \mp \xi_{\rho,x}\xi_{\sigma,x} + 
\eta_{\rho,x} \eta_{\sigma,x}\xi_{\rho,x}\xi_{\sigma,x}].
\end{equation}
This symmetry follows from the relation
$S_R(R^{-1}\ x) \eta_{\alpha,x} S_R(R^{-1}\ x + R^{-1} \hat{\alpha}) = R_{\mu\nu} \eta_{\nu,R^{-1}\ x}$. \\
\noindent (3) {\em Axis Reversal Symmetry:}
\begin{equation}
\psi^a_x \rightarrow (-1)^{x_\rho}\psi^a_{I\ x},
\end{equation}
where $I \equiv I^{(\rho)}$ is the axis reversal $x_\rho \rightarrow -x_\rho$, $x_\tau = x_\tau$, $\tau \neq \rho$.\\
\noindent (4) {\em Global Chiral Symmetry:}
\begin{equation} 
\psi^a_x \rightarrow  (V)^{ab}\psi^b_x, \ x\ \in \ \mbox{even},\ \ 
\psi^a_x \rightarrow  (V^*)^{ab}\psi^b_x,\ x\ \in \mbox{odd}.\ \ 
\end{equation}
where $V$ is an $SU(4)$ matrix in the fundamental representation. Note that the fields at even and odd sites transform differently.

As in the continuum, the above symmetries forbid fermion bilinear mass terms to be generated through radiative corrections. The corresponding mass order parameters were constructed long ago \cite{vandenDoel:1983mf,Golterman:1984cy} and were studied recently in \cite{Catterall:2015zua}. They are given by
\begin{subequations}
\begin{eqnarray}
O^0_{ab}(x) \ &=&\ \psi^a_x\psi^b_x \\
O^1_{\mu,a}(x) \ &=&\ \epsilon_x \xi_{\mu,x}\psi^a_x\ S_\mu\psi^a_{x} \\
O^{2A}_{\mu\nu,a}(x) \ &=&\ \xi_{\mu,x} \xi_{\nu,x+\hat{\mu}} \psi^a_x S_\mu S_\nu\psi^a_{x} \\
O^{2B}_{\mu\nu,a}(x) \ &=&\ \  \epsilon_x\xi_{\mu,x} \xi_{\nu,x+\hat{\mu}} \psi^a_x S_\mu S_\nu\psi^a_{x} \\
O^{3}_{\mu\nu\lambda,a}(x) \ &=&\ \xi_{\mu,x} \xi_{\nu,x+\hat{\mu}} \xi_{\nu,x+\hat{\mu}+\hat{\nu}} \psi^a_x S_\mu S_\nu S_\lambda\psi^a_{x},
\end{eqnarray}
\label{orderparam}
\end{subequations}
where $x$ is a lattice site, $\epsilon_x = (-1)^{x_1+x_2+x_3+x_4}$ and $S_\mu \psi^a_x = \psi^a_{x+\hat{\mu}} + \psi^a_{x-\hat{\mu}}$. Further we assume $\mu\neq\nu\neq \lambda$ in the above expressions. These order parameters naturally vanish in the PMW phase, when fermions are massless. However, we will argue in section \ref{sec4} that even the PMS phase they vanish where fermions become massive.

\section{Fermion Bags, Topology and an Index Theorem}
\label{sec3}

\begin{figure}[t]
\begin{center}
\includegraphics[width=0.5\textwidth]{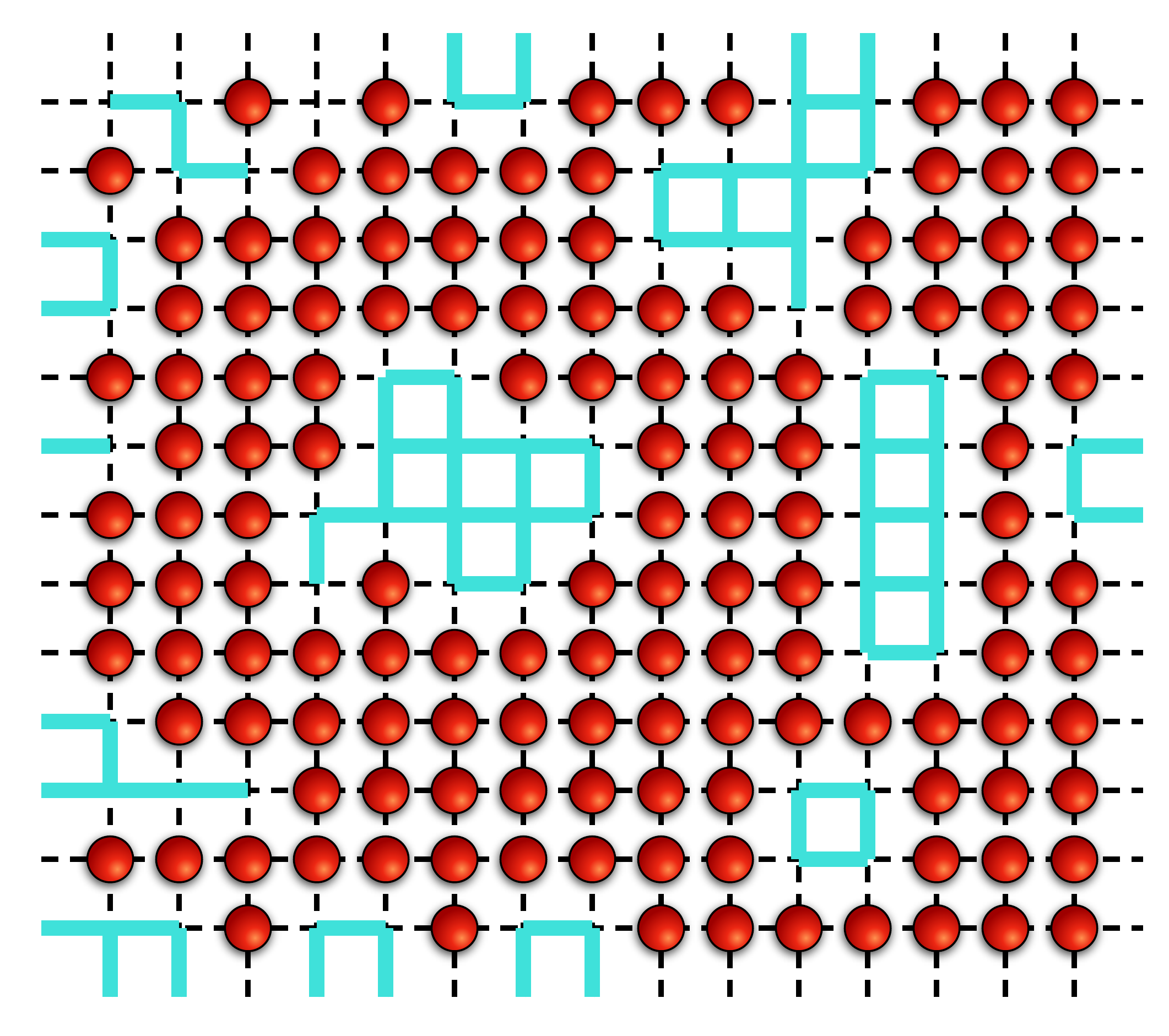}
\end{center}
\caption{\label{fig:1}  An illustration of a fermion bag configuration. The sites with monomers are marked with filled circles. Connected sites without monomers form fermion bags.}
\end{figure}

The partition function of our lattice model whose action is given in (\ref{act}), can be written in the fermion bag approach \cite{PhysRevD.82.025007,Chandraepja13}. In addition to providing an alternate Monte Carlo method to solve lattice fermion field theories, this alternative approach also gives new theoretical insight into the fermion mass generation mechanism involving four-fermion condensates \cite{Chandrasekharan:2014fea}. While the details of this approach was already discussed in \cite{Ayyar:2014eua}, here we repeat the steps for reduced staggered fermions instead of regular staggered fermions. Although the final expression is identical, here it is written in terms of Pfaffians instead of determinants. We first write 
\begin{equation}
Z \ =\ \sum_{[n]}\ \int \prod_x \ [d\psi_x^1 d\psi_x^2 d\psi_x^3 d\psi_x^4]\ 
\ \prod_a \exp\Big(-\frac{1}{2}\ \sum_{x,y} \psi_x^a \ M_{x,y}\ \psi_x^a\Big)\ 
\prod_x\ \Big(U \psi^4_x \psi^3_x \psi^2_x \psi^1_x \Big)^{n_x}.
\label{pf}
\end{equation}
where $[n]$ is a configuration of monomers defined by the binary field $n_x=0,1$ that represents the absence ($n_x=0$) or presence ($n_x=1$) of a monomer (see Fig.~\ref{fig:1} for an illustration). We can perform the Grassmann integral at the sites that contain the monomer first to obtain
\begin{equation}
Z \ =\ \sum_{[n]}\ U^{N_m}\ \prod_a \int [d\psi_{x_1}^a d\psi_{x_2}^a...] 
\ \exp\Big(-\frac{1}{2}\ \sum_{x,y} \psi_x^a \ W_{x,y}\ \psi_y^a\Big),
\end{equation}
where $N_m$ is the total number of monomers in the configuration $[n]$, and sum in the exponent is only over free sites (i.e., sites without monomers) ordered in a convenient way say $x_1,x_2,...$ and $W_{x,y}$ is the reduced staggered Dirac matrix $M_{x,y}$ connecting only the free sites. Performing the remaining Grassmann integration over the free sites we obtain 
\begin{equation}
Z \ =\ \sum_{[n]} U^{N_m} \big(\mathrm{Pf}(W)\big)^4
\label{fbpf}
\end{equation}
where $\mathrm{Pf}(W)$ refers to the Pfaffian of the matrix $W$. Note that the sign of  $\mathrm{Pf}(W)$ is ambiguous and depends on the order in which the free sites are chosen in the definition of the matrix $W$. However, this ambiguity cancels in the full partition function and all physical correlation functions. Since $W$ is an anti-symmetric matrix connecting only even and odd sites, the matrix $W$ can be expressed as a block matrix by separating even and odd sites into separate blocks with non-zero entries only in the off-diagonal block. Hence $\mathrm{Pf}(W)$ is the determinant of this off-diagonal block. When the number of odd and even sites are different then $\mathrm{Pf}(W) = 0$.

Free fermion bags refer to the connected set of free sites which do not belong to a monomer. A monomer configuration can in principle have different disconnected fermion bags, which implies that
\begin{equation}
\mathrm{Pf}(W)\ =\ \prod_{\cal B}\ \mathrm{Pf}(W_{\cal B})
\end{equation}
where the matrix $W_{\cal B}$ refers to the reduced staggered Dirac matrix $W$ connecting only the sites within the bag ${\cal B}$ and the product is over all bags. We will refer to $W_{\cal B}$ as the {\em fermion bag matrix}.  If $S_{\cal B}$ is the total number of sites of the fermion bag, then $W_{\cal B}$ is also an $S_{\cal B} \times S_{\cal B}$ anti-symmetric matrix with non-zero entries only between even and odd sites. It is easy to see that $\mathrm{Pf}(W_{\cal B})=0$ for a bag with an unequal number of even and odd sites.

\begin{figure}[t]
\begin{center}
\includegraphics[width=.5\textwidth]{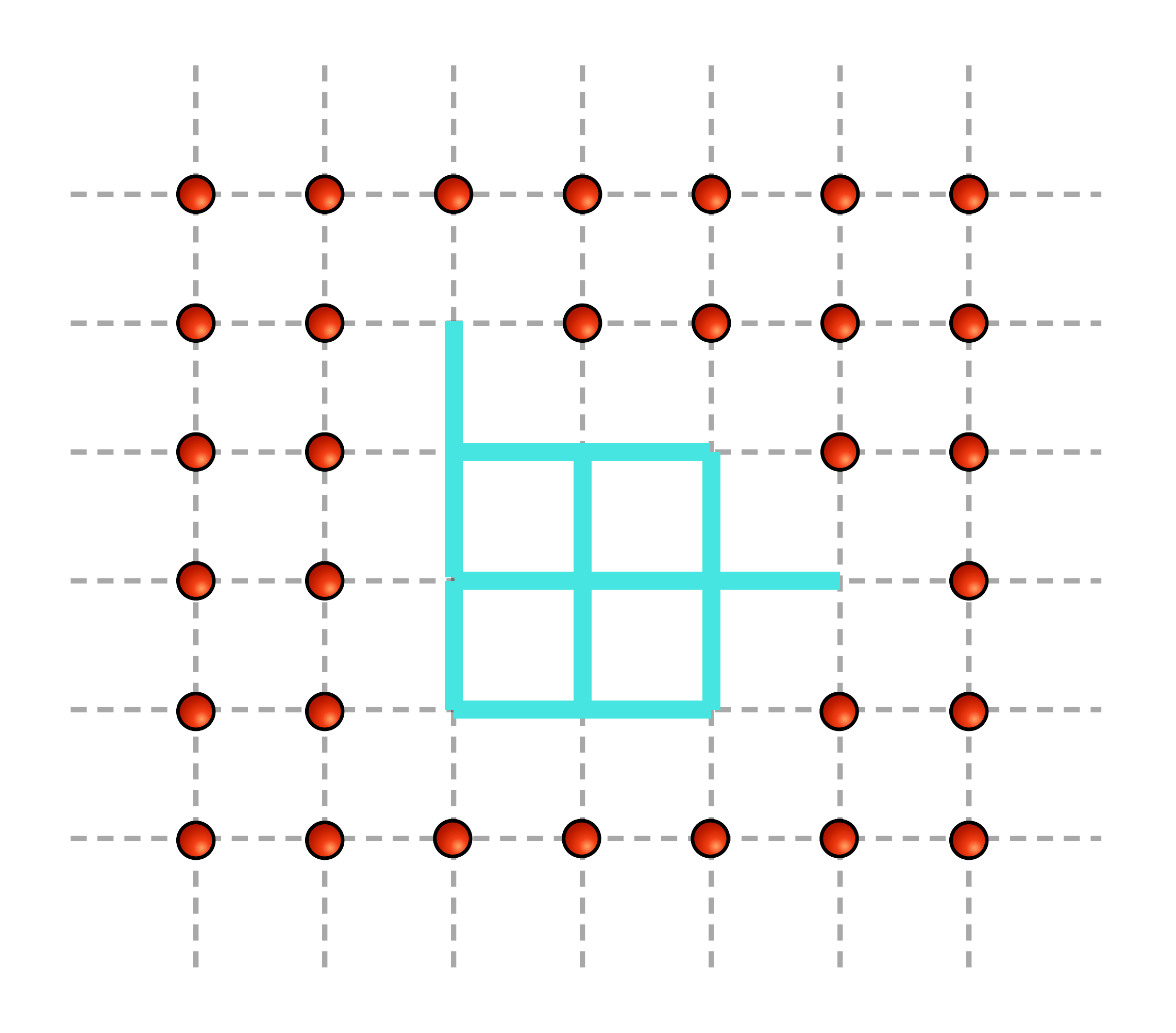}
\end{center}
\caption{\label{fig:2}  An illustration of a $\nu=1$ topological fermion bag configuration. The Dirac matrix $W_{\cal B}$ associated with this bag will contain at least one zero mode. This connection between topology and zero modes is analogous to the index theorem of the massless Dirac operator in non-Abelian gauge theories.}
\end{figure}

Let us now discuss a curious connection between the zero modes of the fermion bag matrix $W_{\cal B}$ and the topology of the fermion bag ${\cal B}$. This connection is analogous to the well known index theorem of the massless Dirac operator in non-Abelian gauge theories \cite{Atiyah:1971rm,Smit:1986fn,Adams:2009eb}. Note that when a bag does not contain an equal number of even and odd sites $W_{\cal B}$ is a matrix with zero modes. If we introduce the concept of a topological charge for the fermion bag through the integer $\nu = n_e - n_o$ (where $n_e$ ($n_o)$ refer the number of even (odd) sites of the bag), then it is easy to argue that the fermion bag matrix $W_{\cal B}$, will have at least $|\nu|$ zero modes, similar to the index of the massless Dirac operator in non-Abelian gauge theories. An example of a $\nu = 1$ topological fermion bag is shown in Fig.~\ref{fig:2}. The analogy with non-Abelian gauge theories extends even further. For example,  in certain massless four-fermion models where a chiral symmetry forbids the presence of the chiral condensate, they can still acquire non-zero expectation values due to the presence of fermion bags with topological charge $\nu = \pm 1$ \cite{Chandrasekharan:2014fea}. This is similar to the fact that chiral condensates obtain a non-zero contribution in QCD with a single massless quark flavor due to the presence of gauge field configurations with topological charge $\nu = \pm 1$ \cite{Leutwyler:1992yt}.

\section{Absence of SSB at Strong Couplings}
\label{sec4}

The conventional wisdom is that when fermions become massive, one or more of the fermion bilinear mass order parameters given in (\ref{orderparam}) acquire a non-zero expectation value due to spontaneous breaking of some of the lattice symmetries. However, it was discovered long ago that the usual single site order parameter vanishes at sufficiently strong couplings even though fermions are massive \cite{Hasenfratz:1988vc,Hasenfratz:1989jr,Lee:1989mi,Bock:1990tv,Bock:1990cx}. More recently the vanishing of all bilinear mass order parameters at strong couplings was studied in \cite{Catterall:2015zua}. In this section we give analytic arguments for this result within the fermion bag approach. Our aim is to illustrate the importance of topology and zero modes of the fermion bag matrix in some of these arguments. A simple extension of these arguments allow us to also conclude the absence of any spontaneous symmetry breaking.

All bilinear mass order parameters given in (\ref{orderparam}) can be written compactly in the form
\begin{equation}
O^\alpha(x) = \sum_y f^\alpha_{a,b}(x,y)\psi^a_x\psi^b_y,
\end{equation}
where $\alpha = 0,1,2A,2B,3$ and $f^\alpha_{a,b}(x,y)$ is appropriately defined with non-zero values only when $x$ and $y$ lie within a hypercube. On a finite lattice we expect $\langle O^\alpha(x)\rangle = 0$ purely from symmetry transformations on Grassmann fields, assuming boundary conditions do not break the symmetries \footnote{our choice of anti-periodic boundary conditions fall in this class}. In the fermion bag approach this vanishing of the symmetry order parameter can be understood through the following three facts:
\begin{enumerate}
\item $\langle O^\alpha(x)\rangle$ can get non-zero contributions from a fermion bag configuration only when both the Grassmann fields in $O^\alpha(x)$ are present within the same fermion bag. To show this let us prove that the weight of a fermion bag vanishes due to the insertion of a single $\psi^a_x$. First note that inserting $\psi^a_x$ in the path integral means that $x$ must be a free site within a fermion bag which we refer to as ${\cal B}_x$. Inserting $\psi^a_x$ and performing the Grassmann integration within the bag gives
\begin{equation}
\mathrm{Pf}(W_{{\cal B}_x}([x]))\ =\ \int\  \prod_{x'\in {\cal B}_x}\ [d\psi^a_{x'}]\ \psi^a_x\ \exp\Big(-\frac{1}{2}\sum_{x,y\in {\cal B'}_x} \psi^a_x W_{x,y} \psi^a_y\Big).
\end{equation}
Note that this is equivalent to removing the site $x$ from the bag and the matrix $W_{{\cal B}_x}([x])$ refers to the fermion bag matrix without the site $x$. Note this matrix has one row and one column less than $W_{{\cal B}_x}$ which contains the site $x$. Without $\psi^a_x$, the above Grassmann integral would give $\mathrm{Pf}(W_{{\cal B}_x})$. Since $x$ will either be an even or an odd site, removing it changes the topology of the fermion bag as defined in the previous section. Thus, if $\mathrm{Pf}(W_{{\cal B}_x}) \neq 0$, then $\mathrm{Pf}(W_{{\cal B}_x}([x])) = 0$ and vice versa. Since the weight of the fermion bag involves a product of four Pfaffians, one for each flavor, the fermion bag weight always vanishes in the presence of a single $\psi^a_x$ source term inside it.

\item When $\alpha=0,2A,2B$, contribution to $\langle O^\alpha(x)\rangle$ from every single fermion bag configuration vanishes because the fermion bag weight that contains the fermion source terms vanishes. In the fermion bag approach these expectation values are given by
\begin{eqnarray}
\langle O^0(x)\rangle &=& \frac{1}{Z} 
\Big\{\sum_{[n]} U^{N_m} \Big(\mathrm{Pf}(W_{{\cal B}_{x}}[x])\Big)^2\Big(\mathrm{Pf}(W_{{\cal B}_{x}})\Big)^2 
\prod_{{\cal B} \neq {\cal B}_{x}} \Big(\mathrm{Pf}(W_{\cal B})\Big)^4\Big\},
\\
\langle O^{2A,2B}(x)\rangle &=& \frac{1}{Z} 
\Big\{\sum_{y} f^{2A,2B}_{a,a}\sum_{[n]} U^{N_m} 
\Big(\mathrm{Pf}(W_{{\cal B}_{x,y}}[x,y])\Big)\Big(\mathrm{Pf}(W_{{\cal B}_{x,y}})\Big)^3
\prod_{{\cal B} \neq {\cal B}_{x}} \Big(\mathrm{Pf}(W_{\cal B})\Big)^4\Big\},
\nonumber \\
\end{eqnarray}
where ${\cal B}_{x}$, $\mathrm{Pf}(W_{{\cal B}_x})$ and $\mathrm{Pf}(W_{{\cal B}_x}([x]))$ were already defined above. We now define ${\cal B}_{x,y}$ as the free fermion bag containing both the sites $x,y$, $\mathrm{Pf}(W_{{\cal B}_{x,y}})$ is the Pfaffian of that fermion bag matrix, and $\mathrm{Pf}(W_{{\cal B}_{x,y}}([x,y]))$ is the Pfaffian of the fermion bag matrix where $x$ and $y$ are also dropped from the bag ${\cal B}_{x,y}$. 
Mathematically,
\begin{equation}
\mathrm{Pf}(W_{{\cal B}_{x,y}}([x,y]))\ =\ \int\  \prod_{x'\in {\cal B}_x}\ [d\psi^a_{x'}]\ \psi^a_x \ \psi^a_y\ \exp\Big(-\frac{1}{2}\sum_{x,y \in {\cal B}_{x,y}} \psi^a_x W_{x,y} \psi^a_y\Big),
\end{equation}
and
\begin{equation}
\mathrm{Pf}(W_{{\cal B}_{x,y}})\ =\ \int\  \prod_{x'\in {\cal B}_x}\ [d\psi^a_{x'}]\ \exp\Big(-\frac{1}{2}\sum_{x,y \in {\cal B}_{x,y}} \psi^a_x W_{x,y} \psi^a_y\Big).
\end{equation}
Since $O^0(x)$ contains $\psi^a_x\psi^b_x$ where $a\neq b$, out of the four Pfaffians contributing to the weight of the fermion bag ${\cal B}_x$, two involve matrices that contain the site $x$ and two involve matrices that do not contain it. Their product vanishes for topological reasons like before, since one of the two matrices will have an extra even or odd site and its pfaffian will vanish. In the case of $O^{2A,2B}(x)$ the weight function $f^{2A,2B}_{a,b}(x,y) \neq 0$ when $a=b$, but when both $x,y$ belong to either even sites or odd sites. Thus, either $W_{{\cal B}_{x,y}}[x,y]$ or $W_{{\cal B}_{x,y}}$ has two extra even or odd sites. Again for topological reasons like before, the pfaffian of one of these two matrices will vanish.

\item The contribution to $\langle O^\alpha(x)\rangle$ for $\alpha = 1,3$ from a given fermion bag configuration may be non-zero, since in these two cases the fermion bag weight that contains the fermion source terms can have a non-zero weight. For these bilinears, since $a=b$ the expression for the expectation value is given by
\begin{equation}
\langle O^{1,3}(x)\rangle = \frac{1}{Z} \sum_y \ f^\alpha_{a,a}(x,y)
\Big\{\sum_{[n]} U^{N_m} 
\Big(\mathrm{Pf}(W_{{\cal B}_{x,y}}[x,y])\Big)\Big(\mathrm{Pf}(W_{{\cal B}_{x,y}})\Big)^3 
\prod_{{\cal B} \neq {\cal B}_{x,y}} \Big(\mathrm{Pf}(W_{\cal B})\Big)^4\Big\}
\end{equation}
where ${\cal B}_{x,y}$, $\mathrm{Pf}(W_{{\cal B}_{x,y}})$ and $\mathrm{Pf}(W_{{\cal B}_{x,y}}([x,y]))$ were defined above. Note that now both $\mathrm{Pf}(W_{{\cal B}_{x,y}})$ and $\mathrm{Pf}(W_{{\cal B}_{x,y}}([x,y]))$ can be non-zero.
\end{enumerate}

Using the above three facts we can understand why  $\langle O^\alpha(x)\rangle = 0$ for all values of $\alpha$. When $\alpha = 0,2A,2B$, contribution from each fermion bag configuration vanishes. However, when $\alpha =1,3$ a further sum over contributions from all symmetry transformations of the fermion bag configuration is necessary to show that the expectation value vanishes. Under these transformations fermion bags transform as a classical extended objects in space-time. For example under a shift in some direction, all monomers and fermion bags get shifted by one lattice unit in that direction. Similarly under rotation by $90^o$ about some axis, the full fermion bag configuration rotates by the same amount.  All such configurations obtained by symmetry transformations will have the same weight in the absence of source insertions. This means $\mathrm{Pf}(W_{\cal B})$ remains the same for all fermion bags ${\cal B} \neq {\cal B}_{x,y}$. On the other hand $\mathrm{Pf}(W_{{\cal B}_{x,y}}([x,y]))$ transforms like $\psi^a_x\psi^a_y$ because of the source insertions and hence will cancel due to the sum over symmetry transformations.

\begin{figure*}[t]
\begin{center}
\includegraphics[width=.4\textwidth]{fig1.pdf}
\hspace{0.5in}
\includegraphics[width=.4\textwidth]{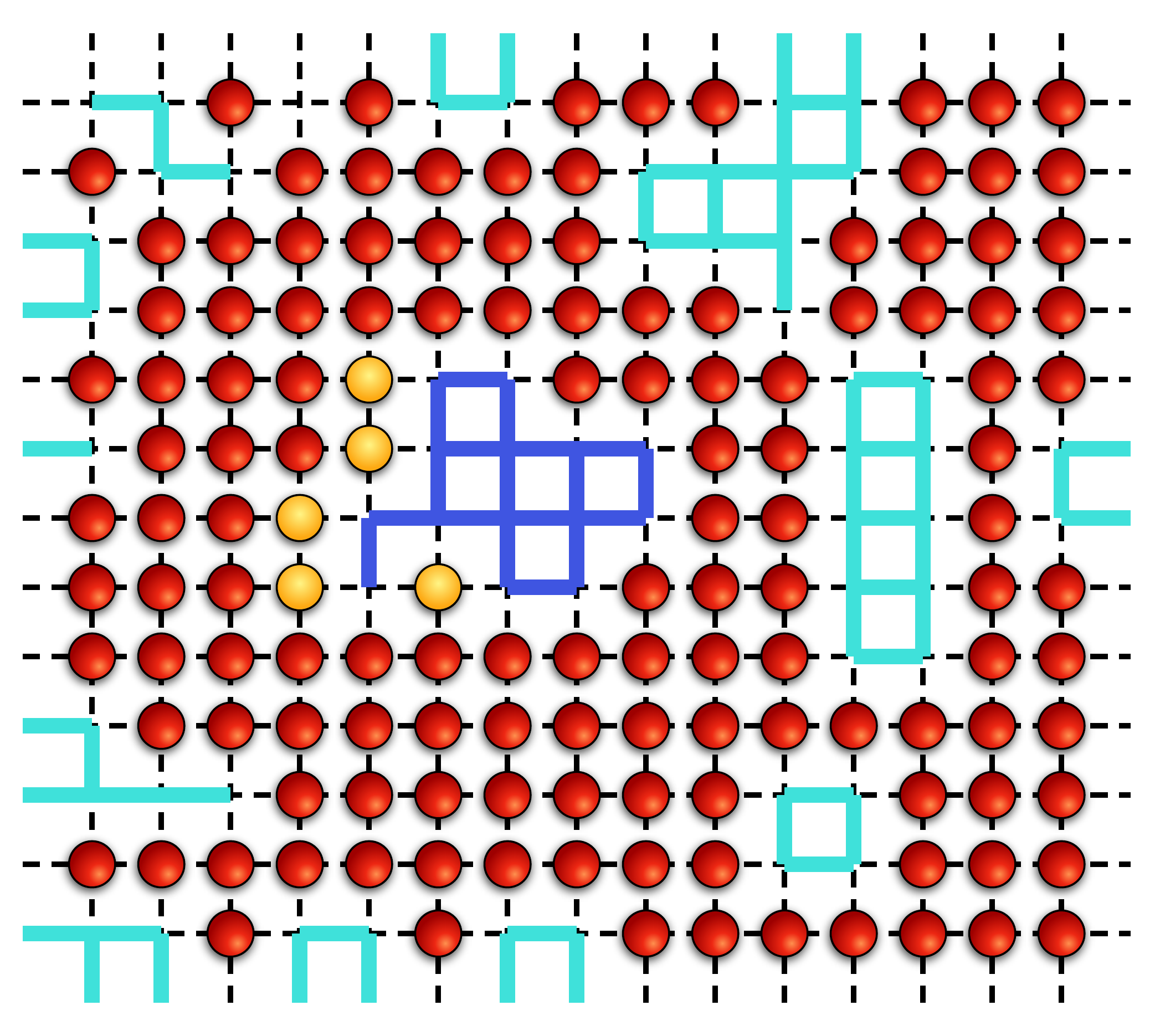}
\end{center}
\caption{\label{fig:fluct}  An illustration of a symmetry fluctuation of a fermion bag when other fermion bags are sufficiently far apart. The fermion bag in the center of the figure on the left has been translated by one unit to the right and shown in the figure on the right. Such a change in a fermion bag is referred to as a symmetry fluctuation and the sites affected during the fluctuation are shown with a different color in the right figure.}
\end{figure*}

Interestingly, if fermion bags are sufficiently far apart symmetry operations can be performed on a single fermion bag without affecting other bags. Such symmetry fluctuations of the single fermion bag that contains the fermion source terms then naturally lead to $\langle O^\alpha(x)\rangle = 0$. Let us illustrate this by considering the calculation of $\langle O^1_{\mu,a}(x)\rangle$ which is given by
\begin{equation}
\langle O^1_{\mu,a}(x)\rangle \ =\ \epsilon(x)\xi_\mu(x) \ \Big(
\langle \psi^a_x \psi^a_{x+\hat{\mu}}\rangle - \langle \psi^a_{x-\hat{\mu}} \psi^a_x\rangle \Big).
\end{equation}
Note that under the shift symmetry we expect 
\begin{equation}
\langle \psi^a_x \psi^a_{x+\hat{\mu}}\rangle = \langle \psi^a_{x-\hat{\mu}} \psi^a_x\rangle.
\end{equation}
which is the reason for $\langle O^1_{\mu,a}(x) \rangle$ to vanish. In the fermion bag approach we have
\begin{equation}
\langle \psi^a_x\psi^a_{x+\hat{\mu}}\rangle = \frac{1}{Z}
\Big\{\sum_{[n]} U^{N_m} \ 
\Big[\mathrm{Pf}(W_{{\cal B}_{x,x+\hat{\mu}}}[x,x+\hat{\mu}]) \Big(\mathrm{Pf}(W_{{\cal B}_{x,x+\hat{\mu}}})\Big)^3\Big] \ \prod_{{\cal B} \neq {\cal B}_{x,x+\hat{\mu}}} \Big(\mathrm{Pf}(W_{\cal B})\Big)^4
\Big\}.
\end{equation}
If the fermion bags are sufficiently far apart, then we can map the bag ${\cal B}_{x-\hat{\mu},x}$ that contributes to $\langle\psi^a_{x-\hat{\mu}}\psi^a_x\rangle$ to another unique bag ${\cal B}^\mu_{x,x+\hat{\mu}}$ obtained by translating ${\cal B}_{x-\hat{\mu},x}$ by one lattice spacing in the $\hat{\mu}$ direction without disturbing any of the other bags. An illustration of such a translation is shown in Fig.~\ref{fig:fluct}. Due to translational invariance we must have
\begin{equation}
\mathrm{Pf}(W_{{\cal B}_{x-\hat{\mu},x}}([x-\hat{\mu},x]))\ =\ 
\mathrm{Pf}(W_{{\cal B}^\mu_{x,x+\hat{\mu}}}([x,x+\hat{\mu}])),
\ \ 
\mathrm{Pf}(W_{{\cal B}_{x-\hat{\mu},x}})\ =\ 
\mathrm{Pf}(W_{{\cal B}^\mu_{x,x+\hat{\mu}}}).
\end{equation}
Since none of the other bags are disturbed, we see that $\langle O^1_\mu(x)\rangle = 0$ simply due the sum over all symmetry fluctuations of the single fermion bag containing the fermion source terms.

The above discussion sheds light on why $\langle O^\alpha(x)\rangle=0$ in a finite system due to symmetry transformations. However, the more important question is to address whether some of these symmetries can break spontaneously. For this one must compute the two point correlation function of local order parameters,
\begin{equation}
C^{\alpha}(x,x') \ =\ \langle O^\alpha(x) O^\alpha(x')\rangle.
\end{equation}
If $C^{\alpha}(x,x') \neq 0$ in the limit when $|x-x'| \rightarrow\infty$, then we say the order parameter is non-zero and the corresponding symmetry is spontaneously broken. Let us now argue that $C^{\alpha}(x,x') = 0$ for all fermion bilinear mass order parameters when $|x-x'| \rightarrow\infty$ at sufficiently strong couplings. We will assume that fermion bags of large size are exponentially suppressed and that fermion bags are well separated from each other so that when fermion bags fluctuate due to symmetry transformations they rarely touch each other. Emperical evidence shows that this assumption is quite reasonable. Hence, contribution from configurations where $x$ and $x'$ lie within the same fermion bag should also be exponentially suppressed in the limit where $|x-x'|$ is large. Thus, a non-zero order parameter requires a non-vanishing contribution from configurations where $x$ and $x'$ are in two different fermion bags. However, we have already argued that $\langle O^\alpha(x)\rangle=0$ within each fermion bag once fluctuations of these two bags are taken into account. Thus, all fermion bilinear mass order parameters must vanish at sufficiently strong coupling, (in the PMS phase) even though fermions are massive. The fact that fermion masses and fermion bilinear condensates need not be related to each other was first presented in \cite{Chandrasekharan:2014fea}.

A straightforward generalization of the above arguments show that any symmetry local order parameter that vanishes within a fermion bag (after taking into account symmetry fluctuations of the fermion bag), cannot develop long range order, as long as fermion bags are well separated from each other and large fermion bags are exponentially suppressed. Since distinct fermion bags are always separated by local singlets (monomers), correlations between them are screened. The only way for long range correlations to arise is due to topology that requires the presence of another fermion bag far away, whose weight vanishes due to zero modes that arise through an index theorem. In such cases while the order parameter vanishes on a finite lattice, two point correlations can develop long range correlations. Examples of lattice models that contain such topological correlations are easy to construct \cite{Chandrasekharan:2014fea}. However, as we have discussed above, our model is different and such topological correlations in symmetry order parameters are absent. Hence there can be no spontaneous symmetry breaking of any lattice symmetries at sufficiently large couplings.

\section{Width of the Intermediate Phase}
\label{sec5}

The arguments of the previous section no longer apply when free fermion bags become large and are not well separated from each other. This occurs in the intermediate coupling region where fermion bilinear condensates can in principle form and lattice symmetries can break spontaneously. As explained in the introduction, it would be exciting to find a 4D lattice model without such an intermediate FM phase, but with a direct PMW-PMS second order phase transition. Unfortunately, earlier studies suggest that our lattice model (\ref{act}) has a {\em wide} intermediate FM phase \cite{Lee:1989mi}, although the phase boundaries were not accurately determined. Using the fermion bag approach, in this section we determine them. Since our algorithms scale badly with system size (especially in 4D), we have been able to perform Monte Carlo calculations only up to $L=12$ (we have one result at $L=14$ at $U=1.75$). Assuming the presence of an intermediate phase and using finite size scaling, we are still able to determine the phase boundaries accurately. In contrast to earlier work, our results point to a surprisingly {\em narrow} intermediate FM phase.

\begin{figure}[t]
\begin{center}
\includegraphics[width=\textwidth]{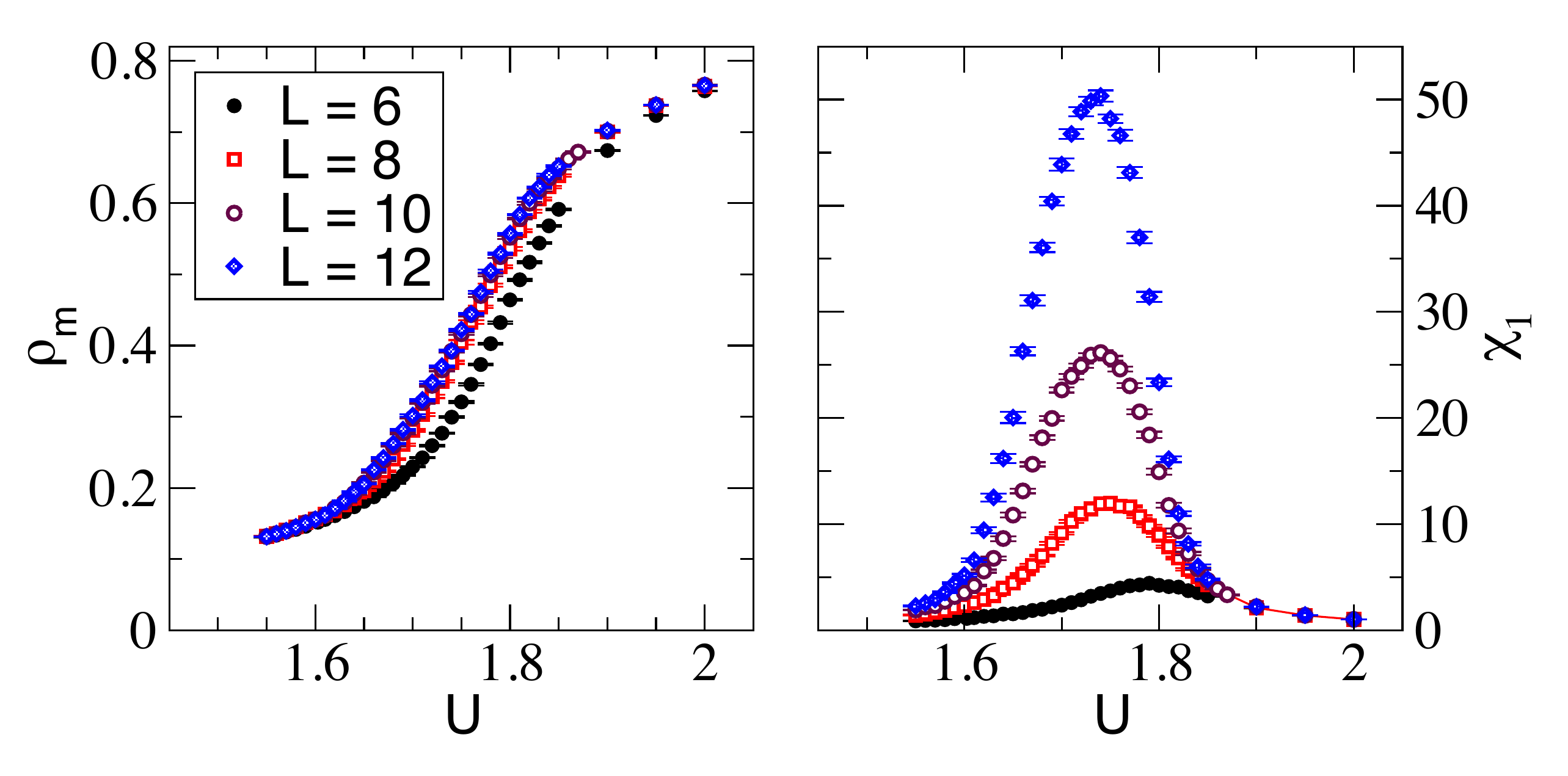}
\end{center}
\caption{\label{fig:rmno} The monomer density (left) and the condensate susceptibility $\chi_1$ (right) plotted as a function of $U$ in the intermediate coupling region for various lattice sizes. There is no sign of a first order transition, but the rapid growth of the susceptibility suggests an intermediate phase with spontaneous breaking of the $SU(4)$ symmetry.}
\end{figure}

We first show results for the four-fermion condensate defined through the monomer density $\rho_m$ in the fermion bag approach using the relation
\begin{equation}
\rho_m = \frac{U}{L^4}\ \sum_x \ \langle \ \psi^4_x\psi^3_x\psi^2_x\psi^1_x \ \rangle.
\end{equation}
Note that with our normalization $\rho_m=0$ at $U=0$ and $\rho_m=1$ at $U=\infty$. In Fig.~\ref{fig:rmno} (on the left side) we plot the behavior of $\rho_m$ as a function of $U$ for various lattice sizes.  The condensate increases rapidly but smoothly between $U=1.5$ and $1.9$ suggesting the absence of any large first order transitions. However, with this data alone it is unclear if there is a single transition due to the absence of an intermediate phase, or two transitions due its presence. For this purpose we compute the two independent susceptibilities 
\begin{equation}
\chi_1 \ =\ \frac{1}{2}\ \sum_{x} \ \langle \psi^1_0\psi^2_0\ \psi^1_x\psi^2_x \rangle,\ \ 
\chi_2 \ =\ \frac{1}{2}\ \sum_{x} \ \langle \psi^1_0\psi^2_0\ \psi^3_x\psi^4_x \rangle,
\end{equation}
that can help in determining if bilinear condensate $\Phi = \langle O^0_{ab}(x) \rangle \neq 0$. In general, $\chi_1 \neq \chi_2$, as can be easily verified for small values of $U$, but for large values of $U$ they become almost similar. Assuming a fermion bilinear condensate forms, the leading behavior at large volumes is expected to scale as $\chi_1 \sim \chi_2 \sim \Phi^2 L^4/4$, since only half the lattice volume contributes in the sum. In other words, a clear signature for the formation of the condensate is the volume scaling of the susceptibilities and that for large $L$ the two susceptibilities become identical.

\begin{figure}[t]
\begin{center}
\includegraphics[width=\textwidth]{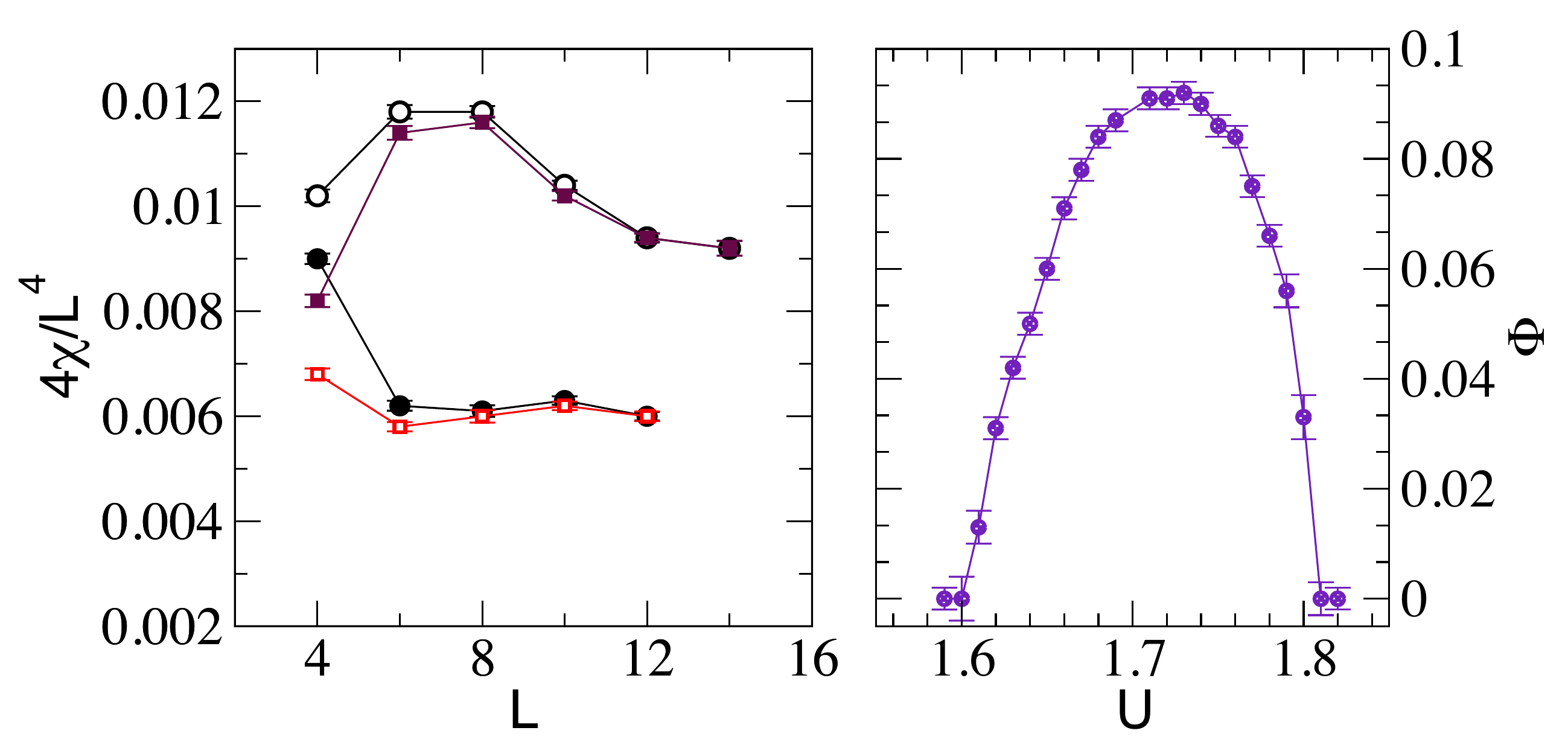}
\end{center}
\caption{\label{fig:susvsL}  The plots on the left show $2\chi_1/L^4$ and $2\chi_2/L^4$ as a function of $L$ at $U=1.67$ (squares) and $1.75$ (circles). Also $\chi_1$ is higher than $\chi_2$. The plot on the right shows the condensate $\Phi = \langle O^0_{ab}(x) \rangle$ as a function of $U$. We see the intermediate FM phase extends roughly from $1.60 \leq U \leq 1.81$.}
\end{figure}

\begin{figure}[t]
\begin{center}
\vbox{
\includegraphics[width=\textwidth]{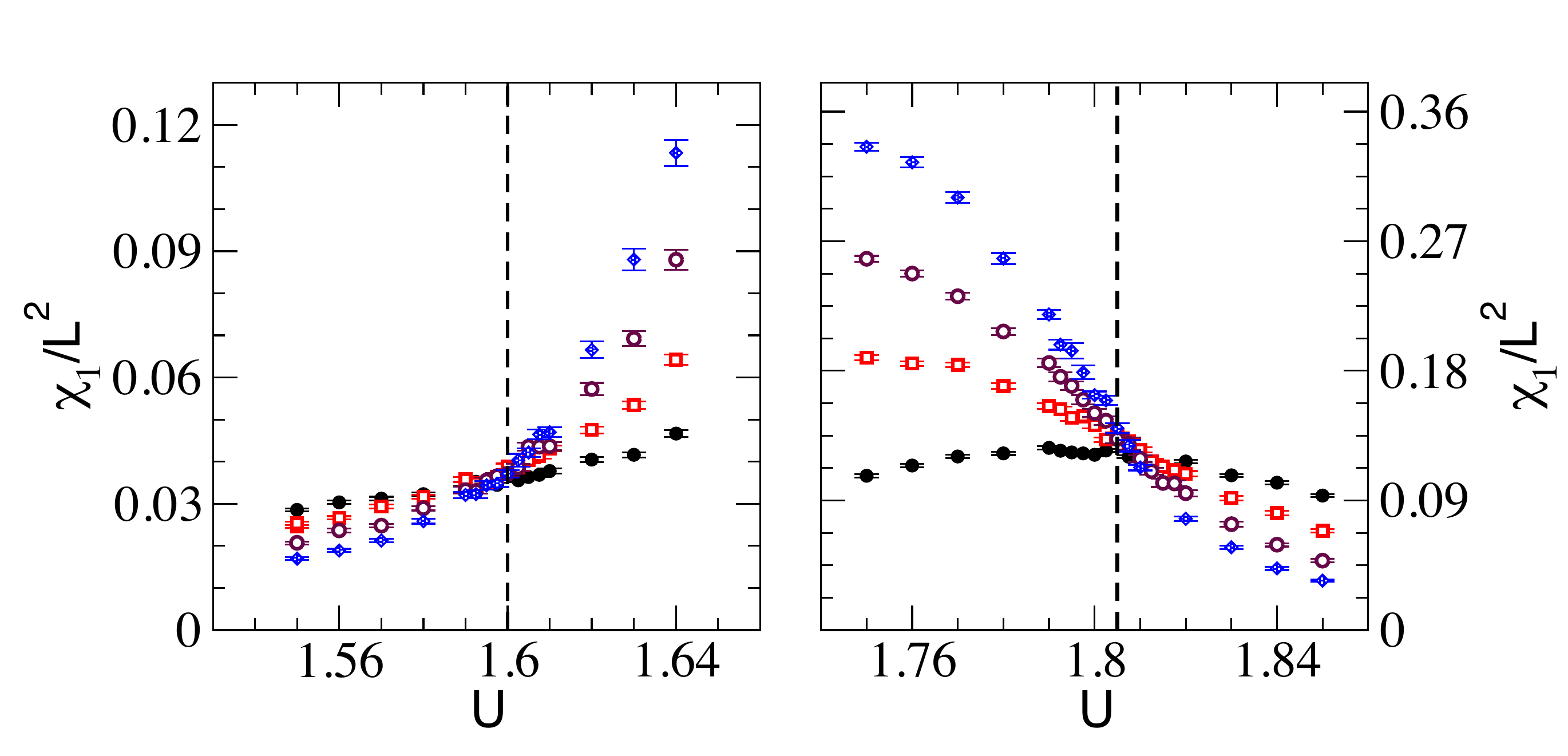}
\includegraphics[width=\textwidth]{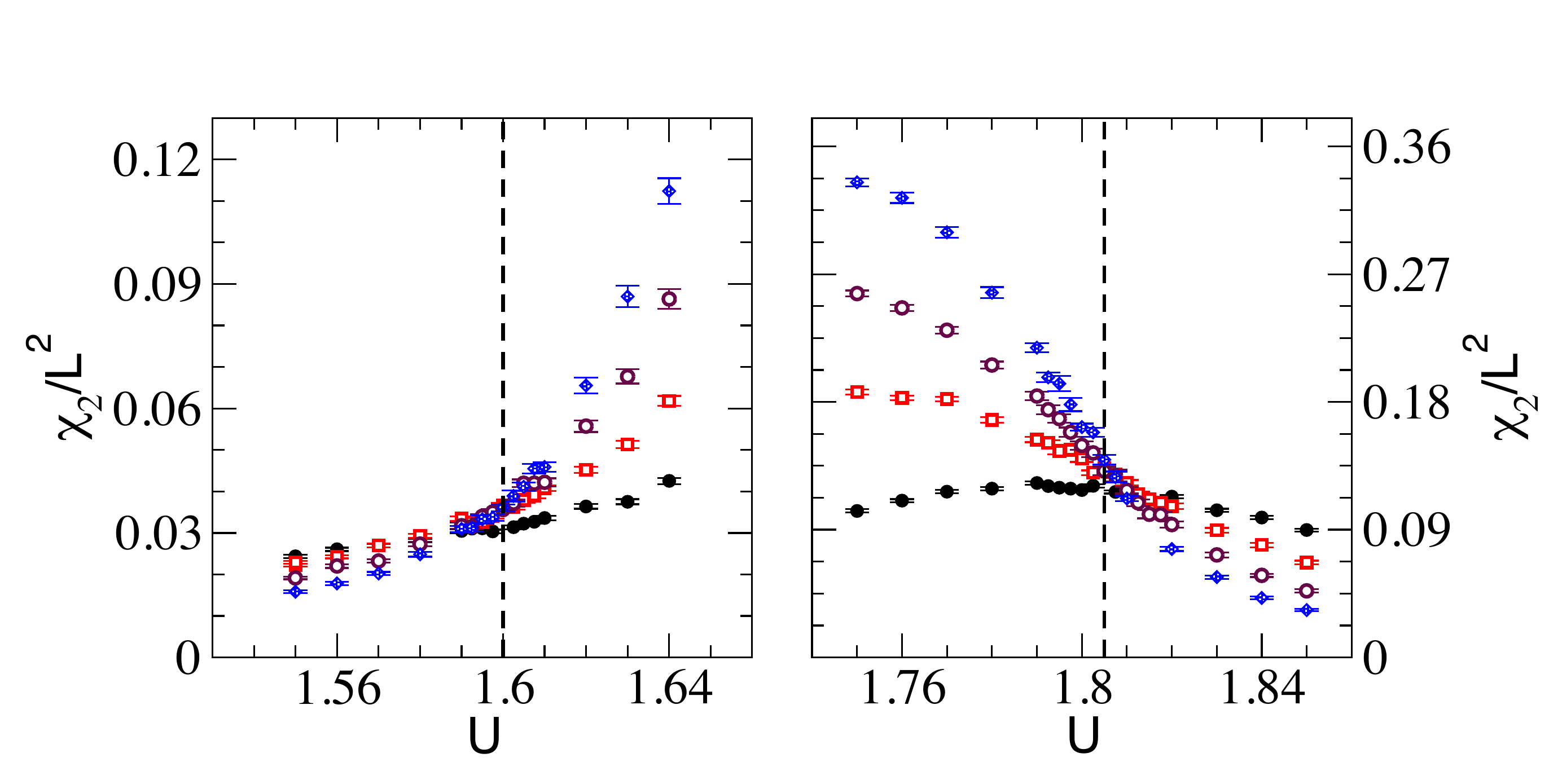}
}
\end{center}
\caption{\label{fig:suscrit}  Plots of $\chi_1/L^2$ (top row) and $\chi_2/L^2$ (bottom row) as a function of $U$ for various lattice sizes near the two transitions. The value of $U$ where the curves cross is shown as the dotted line and indicates the rough location of the critical point.}
\end{figure}

In Fig. (\ref{fig:rmno}) (on the right side) we show the behavior of $\chi_1$ as a function of $U$ for various lattice sizes. For these couplings we find $\chi_2$ to be qualitatively similar. In Fig.\ref{fig:susvsL} (on the left side) we plot both $2\chi_1/L^4$ and $2\chi_2/L^4$ as a function of $L$ at $U=1.67$ and $1.75$. We take the fact that the data seems to be saturating as a sign that a condensate is forming. Further, we observe that $\chi_1 \sim \chi_2$ for the highest two lattices, which provides further evidence for this view point. In contrast, in three dimensions we never found evidence that $\chi_i/L^3$ saturates \cite{Ayyar:2015lrd}. Assuming that the bilinear condensate does form, we fit our data to the form 
\begin{equation}
\chi = \frac{1}{4} \Phi^2 L^4 + b L^2,
\end{equation}
which we found empirically to be a good form for the behavior of the susceptibilities in the intermediate region, to extract the condensate $\Phi = \langle O^0_{ab}(x) \rangle$. This plotted in Fig. \ref{fig:susvsL} (on the right side). 

The fact that $\Phi = \langle O^0_{ab}(x) \rangle \neq 0$ implies that the $SU(4)$ symmetry is broken in the range $1.60 \leq U \leq 1.81$. However, note that this region is much narrower than what was computed in the earlier work. It also means we should have two transitions in our model in quick succession (the PMW-FM transition and the FM-PMS transition). Here we assume that both transitions are second order since we have not seen any reason to believe one of them is first order, but with our small lattice results we cannot rule out the possibility of weak first order transitions. This is especially true for the FM-PMS transition, where the condensate seems to rapidly reducing. Assuming they are second order the PMW-FM transition would follow the Gross Neveu universality while the FM-PMS transition could follow the $SU(4) \sim SO(6)$ spin model universality, both of which would show mean field exponents up to logarithmic corrections. This means
\begin{equation}
\chi_i /L^{2-\eta} \sim f_i((U-U_c) L^{1/\nu})
\end{equation}
where $\eta = 0$ and $\nu = 1/2$ (up to log corrections). In Fig. \ref{fig:suscrit} we plot $\chi_i /L^2 $ versus $U$ for different $L$ values.  As the figure shows, all these curves (for large $L)$  appear to intersect through $U_c$ as expected. We see that $U_c$ for the PMW-FM transition is at roughly $1.60$, and for FM-PMS phase is at around $1.81$, in agreement with our previous conclusion based on computing $\Phi$.

\section{Conclusions}
\label{sec6}

In this work we have studied a lattice field theory model where fermions are massless at weak couplings, but become massive at sufficiently strong couplings even though all fermion bilinear condensates vanish. Fermions seem to acquire their mass through four-fermion condensates. The presence of an intermediate FM phase does not rule out the possibility that this alternate mechanism of mass generation is only a lattice artifact in 4D. On the other hand since the intermediate phase is quite narrow in bare coupling constant space, extending only from $1.60 \leq U \leq 1.81$, it is likely that an extension of the model may reveal the absence of the intermediate phase and may even show the presence of a direct second order PMW-PMS phase transition like in 3D. Such a transition would make the mass generation mechanism through four-fermion condensates interesting even in the continuum.

We can use the continuum model (\ref{contact}) to understand this alternate mechanism of mass generation better. We view four-fermion condensates as a fermion bilinear condensate between a fundamental fermion field and a composite fermion fermion field. For example if $\psi_a(x), a=1,2,3,4$ represents the four components of a Dirac field in four dimensions, then we can view the composite field
\begin{equation}
\bar{\chi}_a(x) = \varepsilon_{abcd} \psi^b(x) \psi^c(x) \psi^d(x),
\end{equation}
as an independent Dirac field such that $\bar{\chi}\psi$ acts as the chirally invariant mass term for a theory that contains both $\psi(x)$ and $\bar{\chi}(x)$. Note that the $U(1)$ fermion number symmetry (\ref{u1fs}) acts as the chiral symmetry for this fermion mass term, while the $U(1)$ chiral symmetry (\ref{u1cs}) acts as the fermion number symmetry of this mass term. Since the continuum model (\ref{contact}) breaks the $U(1)$ fermion number symmetry explicitly, the new type of mass term is always allowed by interactions. However, at weak couplings composite states do not form and the mass term continues to behave as an irrelevant four-fermion coupling. At sufficiently strong couplings, when the composite states form the four-fermion coupling begins to behave like a mass term and becomes relevant.

This fermion mass generation mechanism where fundamental fermions pair with composite fermions is an old idea \cite{Eichten:1985ft,Golterman:1992yha}.  The fact that such a mass generation mechanism can occur without any spontaneous symmetry breaking within a phase (PMS phase) of a regulated microscopic field theory that also contains a phase (PMW phase) with massless fermions was also known before but not emphasized. We find the existence of both these phases within the same regulated microscopic theory exciting, since it means that fermion mass generation can be a dynamical phenomenon purely related to renormalization group arguments rather than symmetry breaking. For all this to be of interest in continuum quantum field theory, there must be a direct second order PMW-PMS transition in the regulated theory. Search for it in 4D would be an interesting research direction for the future.

\acknowledgments

We would like to thank M.~Golterman for pointing us to the lattice literature on the subject. We also thank S.~Catterall, U.-J.~Wiese and C.~Xu for helpful discussions at various stages of this work. SC would like to thank the Center for High Energy Physics at the Indian Institute of Science for hospitality, where part of this work was done. The material presented here is based upon work supported by the U.S. Department of Energy, Office of Science, Nuclear Physics program under Award Number DE-FG02-05ER41368. An important part of the computations performed in this research was done using resources provided by the Open Science Grid, which is supported by the National Science Foundation and the U.S. Department of Energy's Office of Science \cite{Pordes2008,Sfiligoi2009}.

\bibliography{ref,pms,topins}

\providecommand{\href}[2]{#2}\begingroup\raggedright\begin{thebibliography}{10}

\bibitem{Slagle:2014vma}
K.~Slagle, Y.-Z. You, and C.~Xu, {\it {Exotic quantum phase transitions of
  strongly interacting topological insulators}},  {\em Phys.Rev.} {\bf B91}
  (2015), no.~11 115121, [\href{http://arxiv.org/abs/1409.7401}{{\tt
  arXiv:1409.7401}}].

\bibitem{He:2015bda}
Y.-Y. He, H.-Q. Wu, Y.-Z. You, C.~Xu, Z.~Y. Meng, and Z.-Y. Lu, {\it {Bona fide
  interaction-driven topological phase transition in correlated SPT states}},
  \href{http://arxiv.org/abs/1508.06389}{{\tt arXiv:1508.06389}}.

\bibitem{Ayyar:2014eua}
V.~Ayyar and S.~Chandrasekharan, {\it {Massive fermions without fermion
  bilinear condensates}},  {\em Phys. Rev.} {\bf D91} (2015), no.~6 065035,
  [\href{http://arxiv.org/abs/1410.6474}{{\tt arXiv:1410.6474}}].

\bibitem{Catterall:2015zua}
S.~Catterall, {\it {Fermion mass without symmetry breaking}},
  \href{http://arxiv.org/abs/1510.04153}{{\tt arXiv:1510.04153}}.

\bibitem{Ayyar:2015lrd}
V.~Ayyar and S.~Chandrasekharan, {\it {Origin of fermion masses without
  spontaneous symmetry breaking}},  {\em Phys. Rev.} {\bf D93} (2016), no.~8
  081701, [\href{http://arxiv.org/abs/1511.09071}{{\tt arXiv:1511.09071}}].

\bibitem{He:2016sbs}
Y.-Y. He, H.-Q. Wu, Y.-Z. You, C.~Xu, Z.~Y. Meng, and Z.-Y. Lu, {\it {Quantum
  critical point of Dirac fermion mass generation without spontaneous symmetry
  breaking}},  \href{http://arxiv.org/abs/1603.08376}{{\tt arXiv:1603.08376}}.

\bibitem{Stern:1998dy}
J.~Stern, {\it {Two alternatives of spontaneous chiral symmetry breaking in
  QCD}},  \href{http://arxiv.org/abs/hep-ph/9801282}{{\tt hep-ph/9801282}}.

\bibitem{PhysRevD.59.016001}
I.~I. Kogan, A.~Kovner, and M.~Shifman, {\it Chiral symmetry breaking without
  bilinear condensates, unbroken axial ${Z}_{N}$ symmetry, and exact qcd
  inequalities},  {\em Phys. Rev. D} {\bf 59} (Nov, 1998) 016001.

\bibitem{Kanazawa:2015kca}
T.~Kanazawa, {\it {Chiral symmetry breaking with no bilinear condensate
  revisited}},  {\em JHEP} {\bf 10} (2015) 010,
  [\href{http://arxiv.org/abs/1507.06376}{{\tt arXiv:1507.06376}}].

\bibitem{tHooft:1979bh}
G.~'t~Hooft, {\it {Naturalness, chiral symmetry, and spontaneous chiral
  symmetry breaking}},  in {\em {\sl Recent Developments in Gauge Theories},}
  ({'t Hooft, Gerard and Itzykson, C. and Jaffe, A. and Lehmann, H. and Mitter,
  P.K. and Singer, I.M. and Stora, R.}, ed.), vol.~59, p.~135, 1980.

\bibitem{PhysRevLett.45.100}
S.~Coleman and E.~Witten, {\it Chiral-symmetry breakdown in large-n
  chromodynamics},  {\em Phys. Rev. Lett.} {\bf 45} (Jul, 1980) 100--102.

\bibitem{Banks:1991sh}
T.~Banks, {\it {On lattice definitions of chiral gauge theories and the problem
  of anomalies}},  {\em Phys. Lett.} {\bf B272} (1991) 75--80.

\bibitem{Banks:1992af}
T.~Banks and A.~Dabholkar, {\it {Decoupling a fermion whose mass comes from a
  Yukawa coupling: Nonperturbative considerations}},  {\em Phys. Rev.} {\bf
  D46} (1992) 4016--4028, [\href{http://arxiv.org/abs/hep-lat/9204017}{{\tt
  hep-lat/9204017}}].

\bibitem{Hasenfratz:1988vc}
A.~Hasenfratz and T.~Neuhaus, {\it {Nonperturbative Study of the Strongly
  Coupled Scalar Fermion Model}},  {\em Phys.Lett.} {\bf B220} (1989) 435.

\bibitem{Hasenfratz:1989jr}
A.~Hasenfratz, W.-q. Liu, and T.~Neuhaus, {\it {Phase Structure and Critical
  Points in a Scalar Fermion Model}},  {\em Phys.Lett.} {\bf B236} (1990) 339.

\bibitem{Lee:1989mi}
I.-H. Lee, J.~Shigemitsu, and R.~E. Shrock, {\it {Study of Different Lattice
  Formulations of a Yukawa Model With a Real Scalar Field}},  {\em Nucl.Phys.}
  {\bf B334} (1990) 265.

\bibitem{Bock:1990tv}
W.~Bock, A.~K. De, K.~Jansen, J.~Jersak, T.~Neuhaus, and J.~Smit, {\it {Phase
  Diagram of a Lattice SU(2) X SU(2) Scalar Fermion Model With Naive and Wilson
  Fermions}},  {\em Nucl.Phys.} {\bf B344} (1990) 207--237.

\bibitem{Bock:1990cx}
W.~Bock and A.~K. De, {\it {Unquenched Investigation of Fermion Masses in a
  Chiral Fermion Theory on the Lattice}},  {\em Phys.Lett.} {\bf B245} (1990)
  207--212.

\bibitem{Golterman:1990yb}
M.~F.~L. Golterman and D.~N. Petcher, {\it {Decoupling of Doublers and the
  Phase Diagram of Lattice Chiral Fermions for Strong Wilson-yukawa Coupling}},
   {\em Phys. Lett.} {\bf B247} (1990) 370--376.

\bibitem{Gerhold:2007gx}
P.~Gerhold and K.~Jansen, {\it {The Phase structure of a chirally invariant
  lattice Higgs-Yukawa model - numerical simulations}},  {\em JHEP} {\bf 0710}
  (2007) 001.

\bibitem{Bulava:2011jp}
J.~Bulava, P.~Gerhold, G.~W. Hou, K.~Jansen, B.~Knippschild, et~al., {\it
  {Study of the Higgs-Yukawa theory in the strong-Yukawa coupling regime}},
  {\em PoS} {\bf LATTICE2011} (2011) 075.

\bibitem{Shigemitsu:1991tc}
J.~Shigemitsu, {\it {Higgs-Yukawa chiral models}},  {\em Nucl.Phys.Proc.Suppl.}
  {\bf 20} (1991) 515--527.

\bibitem{Eichten:1985ft}
E.~Eichten and J.~Preskill, {\it {Chiral Gauge Theories on the Lattice}},  {\em
  Nucl.Phys.} {\bf B268} (1986) 179.

\bibitem{Golterman:1992yha}
M.~F. Golterman, D.~N. Petcher, and E.~Rivas, {\it {Absence of chiral fermions
  in the Eichten-Preskill model}},  {\em Nucl.Phys.} {\bf B395} (1993)
  596--622, [\href{http://arxiv.org/abs/hep-lat/9206010}{{\tt
  hep-lat/9206010}}].

\bibitem{Alonso:1999hh}
J.~Alonso, P.~Boucaud, V.~Martin-Mayor, and A.~van~der Sijs, {\it {Phase
  diagram and quasiparticles of a lattice SU(2) scalar fermion model in
  (2+1)-dimensions}},  {\em Phys.Rev.} {\bf D61} (2000) 034501.

\bibitem{Senthil:2014ooa}
T.~Senthil, {\it {Symmetry Protected Topological phases of Quantum Matter}},
  {\em Ann. Rev. Condensed Matter Phys.} {\bf 6} (2015) 299,
  [\href{http://arxiv.org/abs/1405.4015}{{\tt arXiv:1405.4015}}].

\bibitem{Callan:1984sa}
C.~G. Callan, Jr. and J.~A. Harvey, {\it {Anomalies and Fermion Zero Modes on
  Strings and Domain Walls}},  {\em Nucl. Phys.} {\bf B250} (1985) 427.

\bibitem{Kaplan:1992bt}
D.~B. Kaplan, {\it {A Method for simulating chiral fermions on the lattice}},
  {\em Phys. Lett.} {\bf B288} (1992) 342--347,
  [\href{http://arxiv.org/abs/hep-lat/9206013}{{\tt hep-lat/9206013}}].

\bibitem{Golterman:1992ub}
M.~F.~L. Golterman, K.~Jansen, and D.~B. Kaplan, {\it {Chern-Simons currents
  and chiral fermions on the lattice}},  {\em Phys. Lett.} {\bf B301} (1993)
  219--223, [\href{http://arxiv.org/abs/hep-lat/9209003}{{\tt
  hep-lat/9209003}}].

\bibitem{Kaplan:1999jn}
D.~B. Kaplan and M.~Schmaltz, {\it {Supersymmetric Yang-Mills theories from
  domain wall fermions}},  {\em Chin. J. Phys.} {\bf 38} (2000) 543--550,
  [\href{http://arxiv.org/abs/hep-lat/0002030}{{\tt hep-lat/0002030}}].

\bibitem{Fidkowski:2009dba}
L.~Fidkowski and A.~Kitaev, {\it {The effects of interactions on the
  topological classification of free fermion systems}},  {\em Phys. Rev.} {\bf
  B81} (2010) 134509, [\href{http://arxiv.org/abs/0904.2197}{{\tt
  arXiv:0904.2197}}].

\bibitem{PhysRevB.83.075103}
L.~Fidkowski and A.~Kitaev, {\it Topological phases of fermions in one
  dimension},  {\em Phys. Rev. B} {\bf 83} (Feb, 2011) 075103.

\bibitem{1367-2630-15-6-065002}
X.-L. Qi, {\it A new class of (2 + 1)-dimensional topological superconductors
  with{$\mathbb {Z}_8$} topological classification},  {\em New Journal of
  Physics} {\bf 15} (2013), no.~6 065002.

\bibitem{Fidkowski:2013jua}
L.~Fidkowski, X.~Chen, and A.~Vishwanath, {\it {Non-Abelian Topological Order
  on the Surface of a 3D Topological Superconductor from an Exactly Solved
  Model}},  {\em Phys.Rev.} {\bf X3} (2013), no.~4 041016,
  [\href{http://arxiv.org/abs/1305.5851}{{\tt arXiv:1305.5851}}].

\bibitem{PhysRevB.92.125104}
T.~Morimoto, A.~Furusaki, and C.~Mudry, {\it Breakdown of the topological
  classification $\mathbb{Z}$ for gapped phases of noninteracting fermions by
  quartic interactions},  {\em Phys. Rev. B} {\bf 92} (Sep, 2015) 125104.

\bibitem{PhysRevB.89.195124}
C.~Wang and T.~Senthil, {\it Interacting fermionic topological
  insulators/superconductors in three dimensions},  {\em Phys. Rev. B} {\bf 89}
  (May, 2014) 195124.

\bibitem{Wen:2013ppa}
X.-G. Wen, {\it {A lattice non-perturbative definition of an SO(10) chiral
  gauge theory and its induced standard model}},  {\em Chin.Phys.Lett.} {\bf
  30} (2013) 111101, [\href{http://arxiv.org/abs/1305.1045}{{\tt
  arXiv:1305.1045}}].

\bibitem{You:2014vea}
Y.-Z. You and C.~Xu, {\it {Interacting Topological Insulator and Emergent Grand
  Unified Theory}},  {\em Phys.Rev.} {\bf B91} (2015), no.~12 125147,
  [\href{http://arxiv.org/abs/1412.4784}{{\tt arXiv:1412.4784}}].

\bibitem{You:2014oaa}
Y.~You, Y.~BenTov, and C.~Xu, {\it {Interacting Topological Superconductors and
  possible Origin of $16n$ Chiral Fermions in the Standard Model}},
  \href{http://arxiv.org/abs/1402.4151}{{\tt arXiv:1402.4151}}.

\bibitem{BenTov:2015gra}
Y.~BenTov and A.~Zee, {\it {The Origin of Families and $SO(18)$ Grand
  Unification}},  \href{http://arxiv.org/abs/1505.04312}{{\tt
  arXiv:1505.04312}}.

\bibitem{PhysRevD.82.025007}
S.~Chandrasekharan, {\it Fermion bag approach to lattice field theories},  {\em
  Phys. Rev. D} {\bf 82} (Jul, 2010) 025007.

\bibitem{Chandraepja13}
S.~Chandrasekharan, {\it Fermion bag approach to fermion sign problems},  {\em
  Eur.~Phys.~J.} {\bf A49} (2013), no.~7 1--12.

\bibitem{Sharatchandra:1981si}
H.~Sharatchandra, H.~Thun, and P.~Weisz, {\it {Susskind Fermions on a Euclidean
  Lattice}},  {\em Nucl.Phys.} {\bf B192} (1981) 205.

\bibitem{Golterman:1984cy}
M.~F. Golterman and J.~Smit, {\it {Selfenergy and Flavor Interpretation of
  Staggered Fermions}},  {\em Nucl.Phys.} {\bf B245} (1984) 61.

\bibitem{vandenDoel:1983mf}
C.~van~den Doel and J.~Smit, {\it {Dynamical Symmetry Breaking in Two Flavor
  SU($N$) and SO($N$) Lattice Gauge Theories}},  {\em Nucl.Phys.} {\bf B228}
  (1983) 122.

\bibitem{Chandrasekharan:2014fea}
S.~Chandrasekharan, {\it {Fermion Bags and A New Origin for a Fermion Mass}},
  {\em PoS} {\bf LATTICE2014} (2014) 309,
  [\href{http://arxiv.org/abs/1412.3532}{{\tt arXiv:1412.3532}}].

\bibitem{Atiyah:1971rm}
M.~F. Atiyah and I.~M. Singer, {\it {The Index of elliptic operators. 5.}},
  {\em Annals Math.} {\bf 93} (1971) 139--149.

\bibitem{Smit:1986fn}
J.~Smit and J.~C. Vink, {\it {Remnants of the Index Theorem on the Lattice}},
  {\em Nucl. Phys.} {\bf B286} (1987) 485--508.

\bibitem{Adams:2009eb}
D.~H. Adams, {\it {Theoretical foundation for the Index Theorem on the lattice
  with staggered fermions}},  {\em Phys. Rev. Lett.} {\bf 104} (2010) 141602,
  [\href{http://arxiv.org/abs/0912.2850}{{\tt arXiv:0912.2850}}].

\bibitem{Leutwyler:1992yt}
H.~Leutwyler and A.~V. Smilga, {\it {Spectrum of Dirac operator and role of
  winding number in QCD}},  {\em Phys. Rev.} {\bf D46} (1992) 5607--5632.

\bibitem{Pordes2008}
R.~Pordes et~al., {\it {The Open Science Grid}},  {\em J. Phys. Conf. Ser.}
  {\bf 78} (2007) 012057.

\bibitem{Sfiligoi2009}
I.~Sfiligoi, D.~C. Bradley, B.~Holzman, P.~Mhashilkar, S.~Padhi, and
  F.~Wurthwein, {\it {he Pilot Way to Grid Resources Using glideinWMS}},  {\em
  WRI World Congress on Computer Science and Information Engineering} {\bf 2}
  (2009) 428--432.

\end{thebibliography}\endgroup

\end{document}